\documentclass[12pt]{article}
\usepackage{graphicx}
\usepackage{cite}

\def \bea{\begin{eqnarray}}
\def \beq{\begin{equation}}

\def \eea{\end{eqnarray}}
\def \eeq{\end{equation}}

\def \3half{\frac{3}{2}}

\begin{document}
\begin{flushright}
TECHNION--PH-2011-15 \\
EFI 11-33 \\
November 2011 \\
\end{flushright}
\centerline{\bf Background dependence of dimuon asymmetry in $\bar p p$
interactions at $\sqrt{s} = 1.96$ TeV}
\bigskip
\centerline{Michael Gronau}
\medskip
\centerline{\it Physics Department, Technion - Israel Institute of Technology}
\centerline{\it Haifa 32000, Israel}
\medskip
\centerline{Jonathan L. Rosner}
\medskip
\centerline{\it Enrico Fermi Institute and Department of Physics,
 University of Chicago} 
\centerline{\it Chicago, IL 60637, U.S.A.} 
\bigskip
\begin{quote}
The D0 Collaboration has reported an anomalous charge asymmetry in the
production of same-sign muon pairs at the Fermilab Tevatron.  The magnitude
of this effect depends on the subtraction of several backgrounds, the most
notable of which is due to kaons being misidentified as muons either through
decays in flight or punch-through.  The present authors suggested a check on
such backgrounds consisting of a tight restriction on the muon impact
parameter $b$, to confirm that this excess was indeed due to $B_{(s)}$ 
meson decays. The D0 Collaboration has performed a related check applying 
transverse impact parameter (IP) restrictions, whose 
implications are discussed.  We study background asymmetry predictions for
events involving two muons with IP bounds which are complementary
to each other. These predictions may be used in
future measurements of the net charge asymmetry from $B_{(s)}$ decays.
\end{quote}

\leftline{\qquad PACS codes:  12.15.Hh, 12.15.Ji, 13.25.Hw, 14.40.Nd} 

\section{Introduction}

Last year the D0 Collaboration reported a charge asymmetry of about 40 times
the standard model value in the production of same-sign muon pairs at the
Fermilab Tevatron \cite{Abazov:2010hv,Abazov:2010hj}.  The magnitude of the
effect depended on the subtraction of several backgrounds, notably muons due to
misidentified or decaying kaons.  
The cross section for $K^-$ on matter is greater than that for $K^+$ on matter, 
so that $K^+$ decays in flight pose a greater source of muons than $K^-$
decays in flight. We suggested that a tight restriction on
the muon impact parameter $b$ could test whether this asymmetry was indeed due
to $B$ meson decays \cite{Gronau:2010cw}.  In an updated version of their
analysis, employing a larger data sample, the D0 Collaboration has implemented
this suggestion, confirming their claim for a larger-than-predicted charge
asymmetry and providing partial separation of $B^0$ and $B_s$ contributions
to the charge asymmetry \cite{Abazov:2011yk}.  The present paper is devoted to
a discussion of their results and their relation to our suggestion.

In Section \ref{sec:ip} we use numbers of 
dimuon events corresponding to different
choices of transverse impact parameter (IP) \cite{Abazov:2011yk} to estimate
the effective value of a related parameter $b > {\rm IP}$.  In Section
\ref{sec:resc} we show that for {\it maximum} values of IP $<(50,80,120)~\mu$m,
scaling the kaon background by a factor of $\simeq$2/3 leads to a vanishing
signal for the dimuon charge asymmetry,
while the net inclusive muon charge asymmetry is shifted away from zero. 
For {\it minimum} values of IP
$>(50,80,120)~\mu$m, with much smaller kaon background, rescaling of that
background does not alter the dimuon 
and inclusive muon asymmetries significantly.
Section \ref{sec:othercuts} discusses predictions of asymmetries for
two muons involving a minimum IP for one and a maximum IP for the other.
We conclude in Section \ref{sec:conc}.

\section{IMPACT PARAMETERS IP AND $b$}
\label{sec:ip}

In their latest study of the like-sign dimuon charge asymmetry at the Tevatron
\cite{Abazov:2011yk}, the D0 Collaboration investigated the dependence of this
asymmetry on the transverse impact parameter IP, a quantity equivalent to what
we called $b_\perp$ \cite{Gronau:2010cw}.  We advocated excluding events with
an impact parameter $b$ greater than a chosen value $b_0$.  If the average
impact parameter in $b$ quark decays is $\langle b \rangle$, the remaining
fraction of dimuons from pairs of hadrons containing $b$ quarks will be
$[1 - \exp(-b_0/\langle b \rangle)]^2$.  Based on a study by the CDF
Collaboration \cite{Aaltonen:2007zza}, we estimated $\langle b \rangle =
350~\mu$m.  For $b_0 = 100~\mu$m, a value we advocated in Ref.\
\cite{Gronau:2010cw}, this quantity is then 0.062.  Such a sample, thus, should
be highly depleted of dimuons from $b$ decays.

Requiring events to involve dimuons with a certain maximum IP ($ \equiv
b_\perp$) is less stringent than requiring them both to have the same maximum
value of $b$, since $b >$ IP.  The two are related by
\beq
b = [{\rm IP}^2 + (b_\parallel \sin \psi)^2 ]^{1/2}~,
\eeq
where $b_\parallel$ and $\psi$ were defined in Ref.\ \cite{Gronau:2010cw}.
It is difficult to impose stringent requirements on $b_\parallel$ \cite{GBPC,
BHPC}, which is why the D0 Collaboration chose instead to restrict the
parameter IP.  With their most stringent constraint, we find their sample of
dimuons from $b$ decays to be reduced by a factor of 6.

The D0 Collaboration studies samples with both muons having IP either less
than or greater than the values 50, 80, and 120 $\mu$m.  We have been
provided \cite{GBPC} with the effect these cuts have on sample sizes.
From this information we are able to extract effective values of $b_0$
which indeed exceed IP, by factors of 2.7 to 3.6.  In what follows we
present details of the calculation.  

The numbers of dimuon events both of whose muons have IP above 50, 80, and 120
$\mu$m are shown in Table \ref{tab:ngt}, while the numbers of events with
IP less than 50, 80, and 120 $\mu$m are shown in Table \ref{tab:nlt}.  The
number of dimuon events due to $b$ decays is
\beq
N^{>>}_{\mu\mu,b} = N^{>>}_{\mu \mu} F^{>>}_{SS} R~,~~
N^{<<}_{\mu\mu,b} = N^{<<}_{\mu \mu} F^{<<}_{SS} R~.
\eeq
where the quantities $F_{SS}$ are given in Tables XXI and XXII of Ref.\
\cite{Abazov:2011yk}, while $R = 94\%$
\cite{GBPC} is the fraction of the ``$SS$'' sample coming from $b$ hadron
decays.  Here we use the superscripts $>>$ and $<<$ to indicate that
both muons have IP either above or below the given value.
The subscript ``$SS$'' \cite{Abazov:2010hv,Abazov:2010hj,Abazov:2011yk} refers
to muons both arising from decays of particles at short distances from the
interaction point.  Muons produced by particles traveling long distances
before decaying in the detector are labeled ``$L$''.

\begin{table}
\caption{Number of dimuon events with both muons having IP above 50, 80, and
120 $\mu$m.
\label{tab:ngt}}
\begin{center}
\begin{tabular}{c c c c} \hline \hline
IP$_{\rm min}$ & $N^{>>}_{\mu \mu}$& $F^{>>}_{SS}$ & $N^{>>}_{\mu\mu,b}$ \\
 \hline
  ($\mu$m)   &  $(10^6)$  &      (\%)      &     $(10^6)$    \\
      50     &    1.680   & 85.63$\pm$3.74 & 1.352$\pm$0.059 \\
      80     &    1.152   & 89.88$\pm$5.10 & 0.973$\pm$0.055 \\
     120     &    0.714   & 91.79$\pm$7.65 & 0.616$\pm$0.051 \\ \hline \hline
\end{tabular}
\end{center}
\end{table}

\begin{table}
\caption{Number of dimuon events with both muons having IP below 50, 80, and
120 $\mu$m.
\label{tab:nlt}}
\begin{center}
\begin{tabular}{c c c c} \hline \hline
IP$_{\rm max}$ & $N^{<<}_{\mu \mu}$& $F^{<<}_{SS}$ & $N^{<<}_{\mu\mu,b}$ \\
 \hline
  ($\mu$m)   &  $(10^6)$  &      (\%)      &     $(10^6)$    \\
      50     &    1.527   & 43.42$\pm$3.75 & 0.623$\pm$0.054 \\
      80     &    2.174   & 48.76$\pm$2.84 & 0.996$\pm$0.058 \\
      120    &    2.857   & 53.66$\pm$2.68 & 1.441$\pm$0.072 \\ \hline \hline
\end{tabular}
\end{center}
\end{table}

The total sample of dimuons due to $b$ hadron decays is $N_{\mu \mu,b} =
N^{<<}_{\mu\mu,b}+N^{>>}_{\mu\mu,b}+N^{<>}_{\mu\mu,b}$,
where the last term is the contribution from events in which one
muon has IP greater than the indicated value and the other has IP less than
the indicated value.  We can calculate this term from the first two:
\beq
N^{<>}_{\mu\mu,b} = 2[N^{<<}_{\mu\mu,b}~N^{>>}_{\mu\mu,b}]^{1/2}~.
\eeq
We should get the same value of $N_{\mu \mu,b}$ for each value of IP; the
values we obtain from the data provided to us are 
(3.81$\pm$0.17, 3.94$\pm$0.16, 3.94$\pm$0.18) $ \times 10^6$ events for 
IP = (50,80,120) $\mu$m.  These are within statistical errors of one another.

The effective values of $b_0$ may now be calculated from the fractions of
dimuon events due to $b$ hadron decays with IP greater than a given amount:
\beq
[\exp(-b_0/\langle b \rangle)]^2 = N^{>>}_{\mu\mu,b}/N_{\mu \mu,b}~,
\eeq
yielding $b_0/\langle b \rangle = (0.518,0.699,0.928)$ for IP = (50,80,120)
$\mu$m.  With $\langle b \rangle = 350~\mu$m as estimated in Ref.\
\cite{Gronau:2010cw}, this yields $b_0 = (181,245,325)~\mu$m, values which
exceed the corresponding ones of IP by factors of (3.6,3.1,2.7).

For the most stringent constraint, taking events with IP $< 50~\mu$m for both
muons, the D0 collaboration is left with $N^{<<}_{\mu\mu,b}/N_{\mu \mu,b} =
0.164 \pm 0.015$, or a sample of about 1/6 the size of that employed 
without imposing bounds on IP values.
We note that while the asymmetry derived in the latter case, $A^b_{sl}=
-0.787\pm 0.172 \pm 0.093$ is 3.9$\sigma$ away from zero~\cite{Abazov:2011yk}, 
a considerably larger asymmetry obtained for IP $<50~\mu$m, 
$A^b_{sl}=-2.779\pm 0.674 \pm 0.694$, is nonzero at 2.9$\sigma$.
The reason for the larger measured asymmetry for muons with low IP may
be that it is driven more by $B_s$ decays than by $B^0$
decays~\cite{Gronau:2010cw}. 

\section{RESCALING THE KAON BACKGROUND}
\label{sec:resc}

In this Section we examine whether rescaling the kaon 
charge asymmetry background from the
value determined in Ref.\ \cite{Abazov:2011yk} can lead to a vanishing
charge asymmetry for same-sign muon pairs
originating from $B_{(s)}$ decays.
We will also discuss the implication of such rescaling on the inclusive muon 
charge asymmetry.

Using the notations of Ref.\ \cite{Abazov:2011yk}, the measured like-sign dimuon 
asymmetry $A$ is related to the  charge asymmetry from $B_{(s)}$ decays 
$A^b_{sl}$  through 
\bea\label{A}
A & = & (F_{SS}C_b + F_{SL}c_b)\,A^b_{sl} + F_KA_K + F_\pi A_\pi + F_p A_p 
+(2-F_{\rm bkg})\Delta~
\nonumber\\
& \equiv & (F_{SS}C_b + F_{SL}c_b)\,A^b_{sl} + A_{\rm bkg}~.
\eea 
Here $F_{SS}C_b + F_{SL}c_b$ is an effective dilution factor
depending on fractions of dimuon events with two short ($S$) muons and with one
short and one long ($L$) muon, while $F_x$ and $A_x$ 
($x=K, \pi, p$) are fractions and asymmetries of muons produced by kaons, pions 
and protons, respectively.
The term $(2-F_{\rm bkg}) \Delta$~ 
[$F_{\rm bkg}\equiv F_K+F_\pi+F_p$, 
$\Delta =(-0.132\pm 0.019)\%$] represents a 
contribution from muon track reconstruction asymmetry.

\begin{table}
\caption{Contributions of background sources to charge asymmetry in like-sign
dimuon sample for nominal analysis of Ref.\ \cite{Abazov:2011yk}.
\label{tab:tab12}}
\begin{center}
\begin{tabular}{l r} \hline \hline
Source                                 & Asymmetry \\ \hline
$F_K A_K \times 10^2$                  & $+0.633 \pm 0.031$ \\
$F_\pi A_\pi \times 10^2$              & $-0.002 \pm 0.023$ \\
$F_p A_p \times 10^2$                  & $-0.016 \pm 0.019$ \\
$(2 - F_{\rm bkg}) \Delta \times 10^2$ & $-0.212 \pm 0.030$ \\ \hline
$A_{\rm bkg} \times 10^2$              & $+0.402 \pm 0.053$ \\ \hline
$A \times 10^2$                        & $+0.126 \pm 0.041$ \\
$(A - A_{\rm bkg}) \times 10^2$        & $-0.276 \pm 0.067$ \\ \hline \hline
\end{tabular}
\end{center}
\end{table}

We begin by quoting in Table \ref{tab:tab12} some relevant entries from Table
XII of Ref.\ \cite{Abazov:2011yk} for like-sign dimuon events, which are the
main source of statistical weight supporting the claim for a charge asymmetry
$A^b_{sl}$.  In Table \ref{tab:tab12} the dominant contribution to
$A_{\rm bkg}$ is from the charge asymmetry in kaon tracks.  The final result
for the dimuon charge asymmetry is given by $(A - A_{\rm bkg})$ divided
by an effective dilution factor, $F_{SS}C_b + F_{SL}c_b = 0.342 \pm 0.028$, or
\beq
A^b_{sl} = (-0.808 \pm 0.202~({\rm stat}) \pm 0.222~({\rm syst}))\%.
\eeq
Thus a reduction of the kaon background by a factor of 0.56 to 0.356 would be
required to achieve a vanishing value of $A - A_{\rm bkg}$ or of $A^b_{sl}$.

Ref.\ \cite{Abazov:2011yk} considered a set of variations on their nominal
analysis entailing the restrictions IP $< (50,80,120)~\mu$m and IP $>
(50,80,120)~\mu$m.  We are grateful to the D0 Collaboration for sharing
the respective entries in Table \ref{tab:tab12} corresponding to each
of these criteria~\cite{D0web}.  The results shown in Tables \ref{tab:lt} and
\ref{tab:gt} 
were obtained by summing products such as $F_KA_K$ over muon $p_T$ bins.

For comparison, we quote in Tables \ref{tab:ltave} and \ref{tab:gtave} the
dominant contributions $F_KA_K$ and $(2-F_{\rm bkg})\Delta$ using averaged
values of $A_K$ and $F_x$ from Tables VII and XXI, XXII in Ref.\
\cite{Abazov:2011yk}, respectively. We note the insignificant differences
between values of background asymmetries in Tables \ref{tab:lt} and
\ref{tab:ltave} and between those in Table \ref{tab:gt} and \ref{tab:gtave}.
Thus, in the next section, where we discuss charge asymmetries in dimuon events
with other IP constraints, we will use averaged values of fractions and
asymmetries rather than summing their products over muon $p_T$ bins. 

\begin{table}
\caption{Contributions of background sources to charge asymmetry in like-sign
dimuon sample for analysis of Ref.\ \cite{Abazov:2011yk} with the restriction
IP $< (50,80,120)~\mu$m.
\label{tab:lt}}
\begin{center}
\begin{tabular}{l r r r} \hline \hline
Source                                &\multicolumn{3}{c}{Asymmetry} \\
                 & IP $< 50~\mu$m & IP $< 80~\mu$m & IP $< 120~\mu$m \\ \hline
$F_K A_K \times 10^2$                 & $+1.421 \pm 0.066$ & $+1.203 \pm 0.053$
 & $+1.047 \pm 0.051$ \\
$F_\pi A_\pi \times 10^2$             & $+0.016 \pm 0.051$ & $-0.001 \pm 0.002$
 & $-0.003 \pm 0.039$ \\
$F_p A_p \times 10^2$                 & $-0.028 \pm 0.041$ & $-0.029 \pm 0.035$
 & $-0.027 \pm 0.032$ \\
$(2 - F_{\rm bkg}) \Delta \times 10^2$ &$-0.166 \pm 0.021$ & $-0.179 \pm 0.023$
 & $-0.188 \pm 0.025$ \\ \hline
$A_{\rm bkg} \times 10^2$             & $+1.243 \pm 0.096$ & $+0.994 \pm 0.082$
 & $+0.829 \pm 0.077$ \\ \hline
$A \times 10^2$                       & $+0.715 \pm 0.083$ & $+0.683 \pm 0.069$
 & $+0.555 \pm 0.060$ \\
$(A - A_{\rm bkg}) \times 10^2$       & $-0.527 \pm 0.127$ & $-0.311 \pm 0.107$
 & $-0.274 \pm 0.098$ \\ \hline \hline
\end{tabular}
\end{center}
\end{table}
\begin{table}
\caption{Contributions of background sources to charge asymmetry in like-sign
dimuon sample for analysis of Ref.\ \cite{Abazov:2011yk} with the restriction
IP $> (50,80,120)~\mu$m.
\label{tab:gt}}
\begin{center}
\begin{tabular}{l r r r} \hline \hline
Source                                &\multicolumn{3}{c}{Asymmetry} \\
                 & IP $> 50~\mu$m & IP $> 80~\mu$m & IP $> 120~\mu$m \\ \hline
$F_K A_K \times 10^2$                  & $+0.205 \pm 0.060$ 
 & $+0.104 \pm 0.074$ & $+0.113 \pm 0.087$ \\
$F_\pi A_\pi \times 10^2$              & $+0.001 \pm 0.007$
 & $-0.002 \pm 0.003$ & $-0.001 \pm 0.002$ \\
$F_p A_p \times 10^2$                  & $-0.005 \pm 0.005$
 & $-0.004 \pm 0.002$ & $-0.002 \pm 0.002$ \\
$(2 - F_{\rm bkg}) \Delta \times 10^2$ & $-0.244 \pm 0.035$
 & $-0.236 \pm 0.036$ & $-0.237 \pm 0.037$ \\ \hline
$A_{\rm bkg} \times 10^2$              & $-0.043 \pm 0.071$
 & $-0.139 \pm 0.083$ & $-0.127 \pm 0.093$ \\ \hline
$A \times 10^2$                        & $-0.302 \pm 0.079$
 & $-0.386 \pm 0.094$ & $-0.529 \pm 0.120$ \\
$(A - A_{\rm bkg}) \times 10^2$        & $-0.259 \pm 0.106$
 & $-0.247 \pm 0.125$ & $-0.402 \pm 0.152$ \\ \hline \hline
\end{tabular}
\end{center}
\end{table}
\begin{table}
\caption{Dominant contributions of background sources to charge asymmetry in like-sign
dimuon sample with the restriction
IP $< (50,80,120)~\mu$m. We use average values of $A_K$ and $F_x$.
\label{tab:ltave}}
\begin{center}
\begin{tabular}{l r r r} \hline \hline
Source                                &\multicolumn{3}{c}{Asymmetry} \\
                 & IP $< 50~\mu$m & IP $< 80~\mu$m & IP $< 120~\mu$m \\ \hline
$F_K A_K \times 10^2$                 & $+1.37 \pm 0.05$ & $+1.16 \pm 0.04$
 & $+1.05 \pm 0.04$ \\
$(2 - F_{\rm bkg}) \Delta \times 10^2$ &$-0.158 \pm 0.023$ & $-0.173 \pm 0.025$
 & $-0.181 \pm 0.026$ \\ \hline \hline
\end{tabular}
\end{center}
\end{table}
\begin{table}
\caption{Dominant contributions of background sources to charge asymmetry in like-sign
dimuon sample with the restriction IP $> (50,80,120)~\mu$m. We use average values of 
$A_K$ and $F_x$.
\label{tab:gtave}}
\begin{center}
\begin{tabular}{l r r r} \hline \hline
Source            &\multicolumn{3}{c}{Asymmetry} \\
                 & IP $> 50~\mu$m & IP $> 80~\mu$m & IP $> 120~\mu$m \\ \hline
$F_K A_K \times 10^2$                  & $+0.31 \pm 0.09$ 
 & $+0.23 \pm 0.13$ & $+0.22 \pm 0.20$ \\
$(2 - F_{\rm bkg}) \Delta \times 10^2$ & $-0.243 \pm 0.035$
 & $-0.249 \pm 0.036$ & $-0.252 \pm 0.037$ \\ \hline \hline
\end{tabular}
\end{center}
\end{table}

If we wish to rescale $F_K A_K$ by a common factor $\lambda$, we may
parametrize
\beq \label{eqn:resc}
A - A_{\rm bkg} = A_0 - \lambda F_K A_K~,
\eeq
with the values of $A_0 \equiv A - [F_\pi A_\pi + F_p A_p + 
(2 - F_{\rm bkg})\Delta]$ and $F_K A_K$ given in Table \ref{tab:resc}.

\begin{table}
\caption{Parameters in Eq.\ (\ref{eqn:resc}) for rescaling contribution of
kaon background.  The nominal parameters \cite{Abazov:2011yk} involve only a
very weak restriction on impact parameter (IP).
\label{tab:resc}}
\begin{center}
\begin{tabular}{c c c} \hline \hline
Criterion       & $A_0 \times 10^2$ & $F_K A_K \times 10^2$ \\ \hline
Nominal         & $+0.356 \pm 0.059$ & $0.633 \pm 0.031$ \\
IP $< 50~\mu$m  & $+0.893 \pm 0.108$ & $1.421 \pm 0.066$ \\ 
IP $< 80~\mu$m  & $+0.892 \pm 0.081$ & $1.203 \pm 0.053$ \\
IP $< 120~\mu$m & $+0.773 \pm 0.082$ & $1.047 \pm 0.051$ \\ \hline
IP $> 50~\mu$m  & $-0.054 \pm 0.087$ & $0.205 \pm 0.060$ \\
IP $> 80~\mu$m  & $-0.144 \pm 0.101$ & $0.104 \pm 0.074$ \\
IP $> 120~\mu$m & $-0.289 \pm 0.126$ & $0.113 \pm 0.087$ \\ \hline \hline
\end{tabular}
\end{center}
\end{table}

\begin{figure}
\begin{center}
\includegraphics[width=0.6\textwidth]{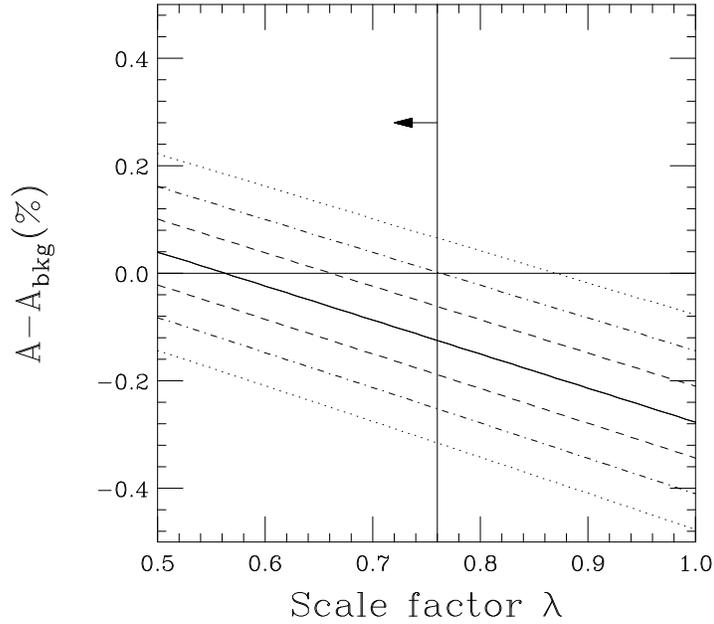}
\end{center}
\caption{Dependence of net dimuon asymmetry $A - A_{\rm bg}$ on scale factor
$\lambda$ for the nominal analysis of Ref.\ \cite{Abazov:2011yk}.  Solid,
dashed, dot-dashed, and dotted sloping lines denote central, $\pm 1 \sigma$,
$\pm 2 \sigma$, and $\pm 3$ sigma values, respectively.  Values of $\lambda$ to
left of vertical line give net asymmetries differing from zero (horizontal
line) by less than $2 \sigma$.  $A - A_{\rm bg}$ is within $2 \sigma$ of zero
for $\lambda < 0.76$. \label{fig:nocut}}
\end{figure}

\begin{figure} \label{fig:50}
\includegraphics[width=0.48\textwidth]{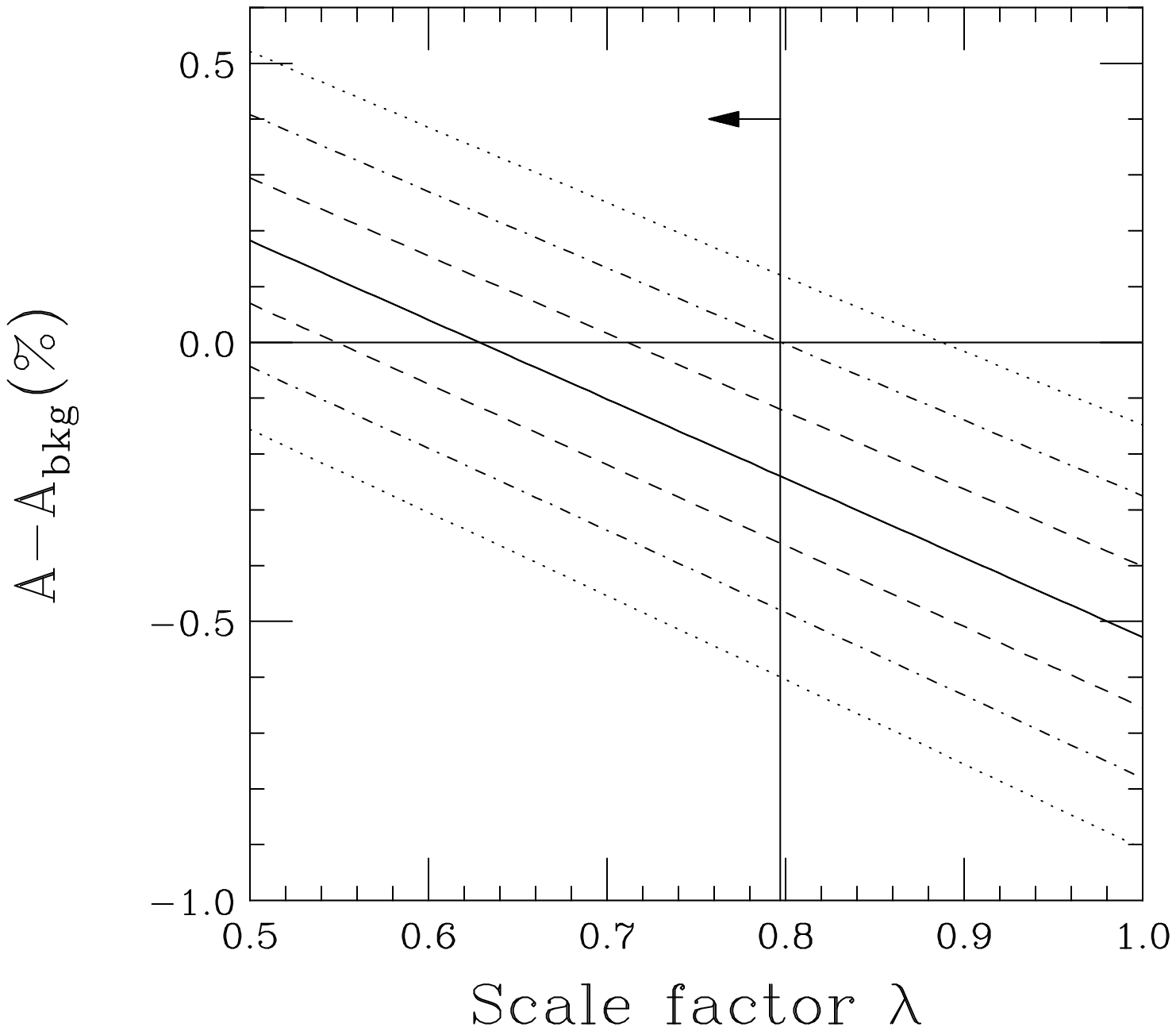}
\includegraphics[width=0.48\textwidth]{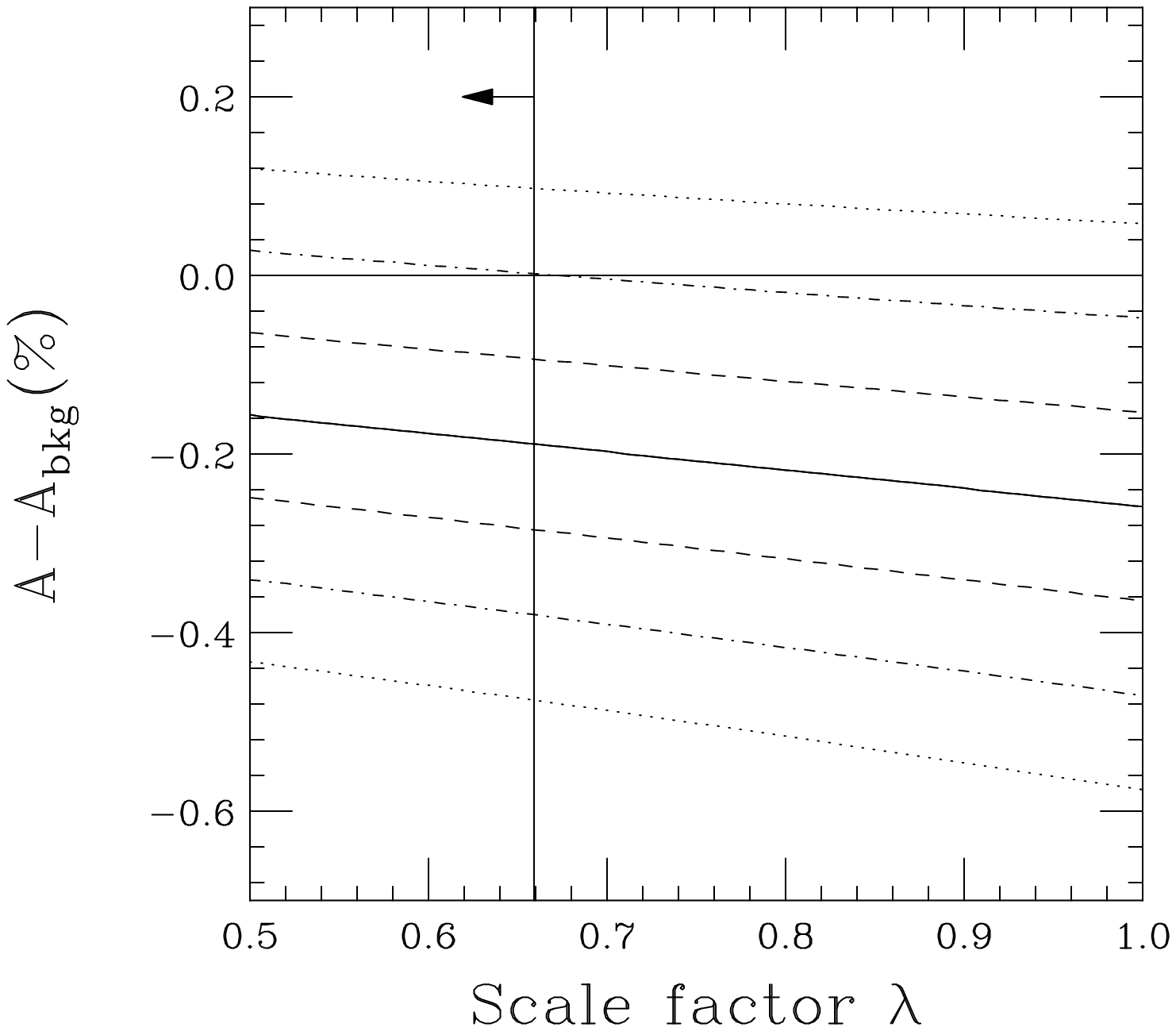}
\caption{Dependence of net dimuon asymmetry $A - A_{\rm bg}$ on scale factor
$\lambda$ for (left) IP $< 50~\mu$m; (right) IP $> 50~\mu$m.  Lines as in
Fig.\ \ref{fig:nocut}.  $A - A_{\rm bg}$ is within $2 \sigma$ of zero for
$\lambda < (0.80,0.66)$.}
\end{figure}

\begin{figure} \label{fig:80}
\includegraphics[width=0.48\textwidth]{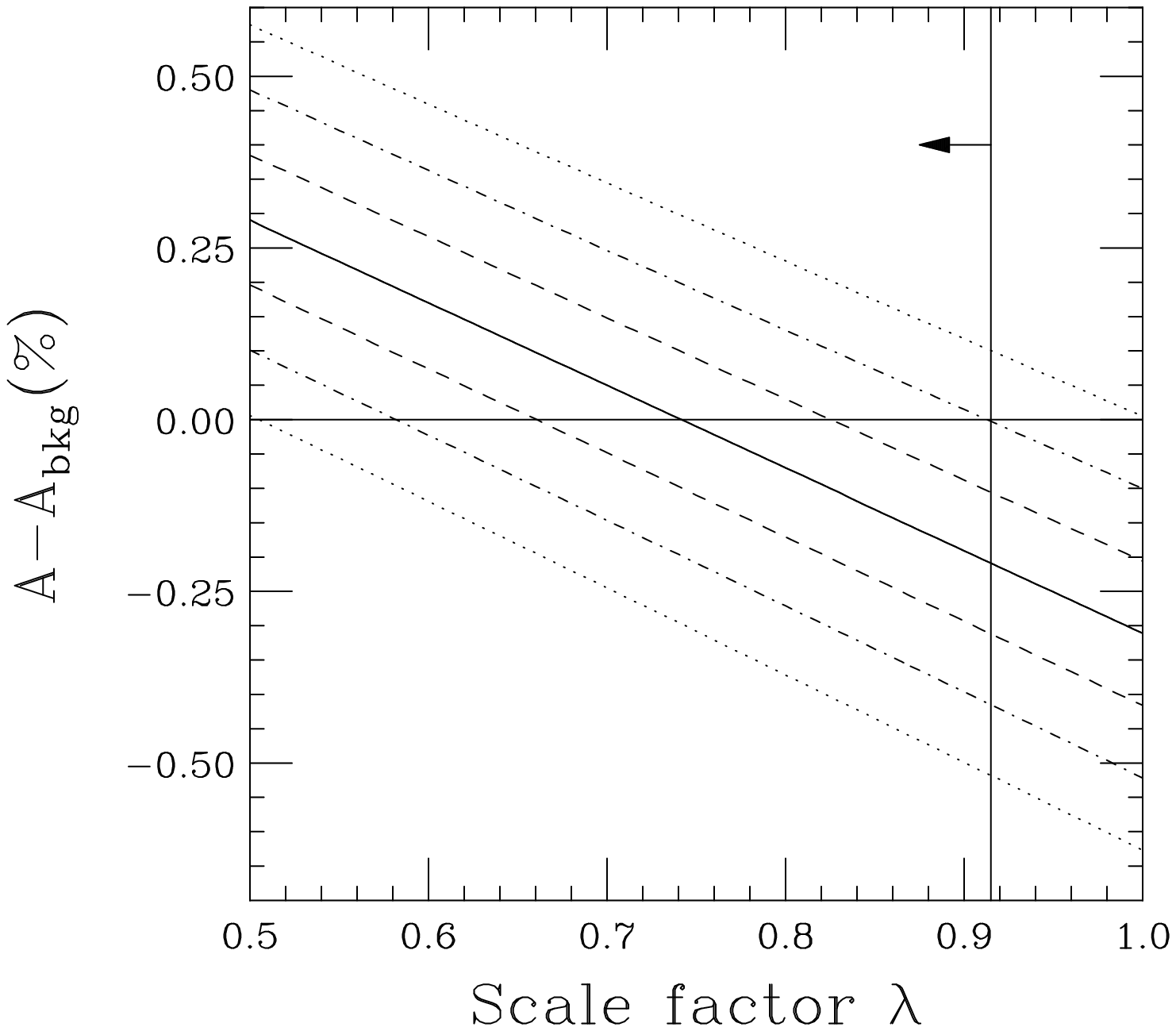}
\includegraphics[width=0.48\textwidth]{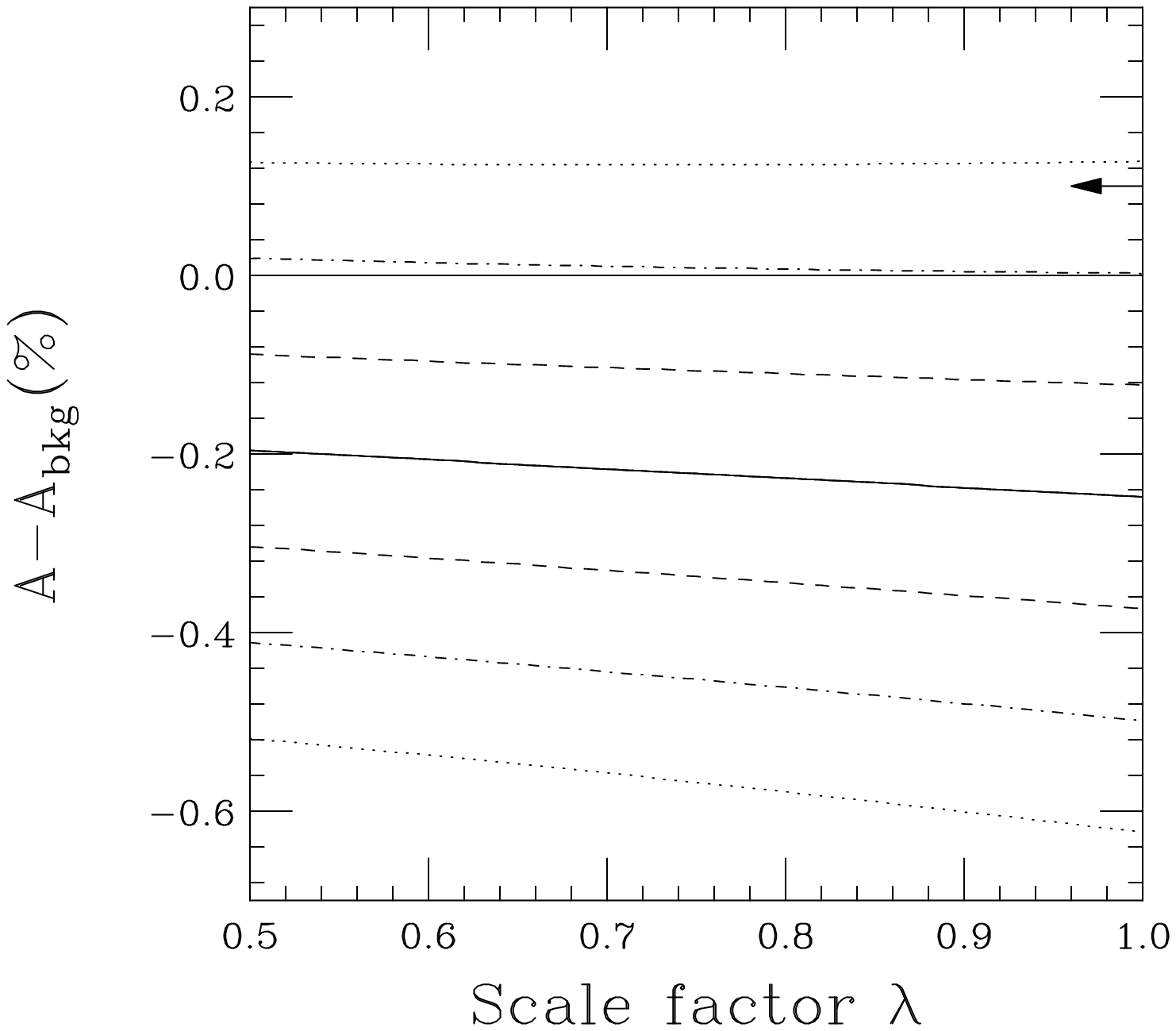}
\caption{Dependence of net dimuon asymmetry $A - A_{\rm bg}$ on scale factor
$\lambda$ for (left) IP $< 80~\mu$m; (right) IP $> 80~\mu$m.  Lines as in
Fig.\ \ref{fig:nocut}.  $A - A_{\rm bg}$ is within $2 \sigma$ of zero for
$\lambda < (0.92,1.00)$.}
\end{figure}

\begin{figure} \label{fig:120}
\includegraphics[width=0.48\textwidth]{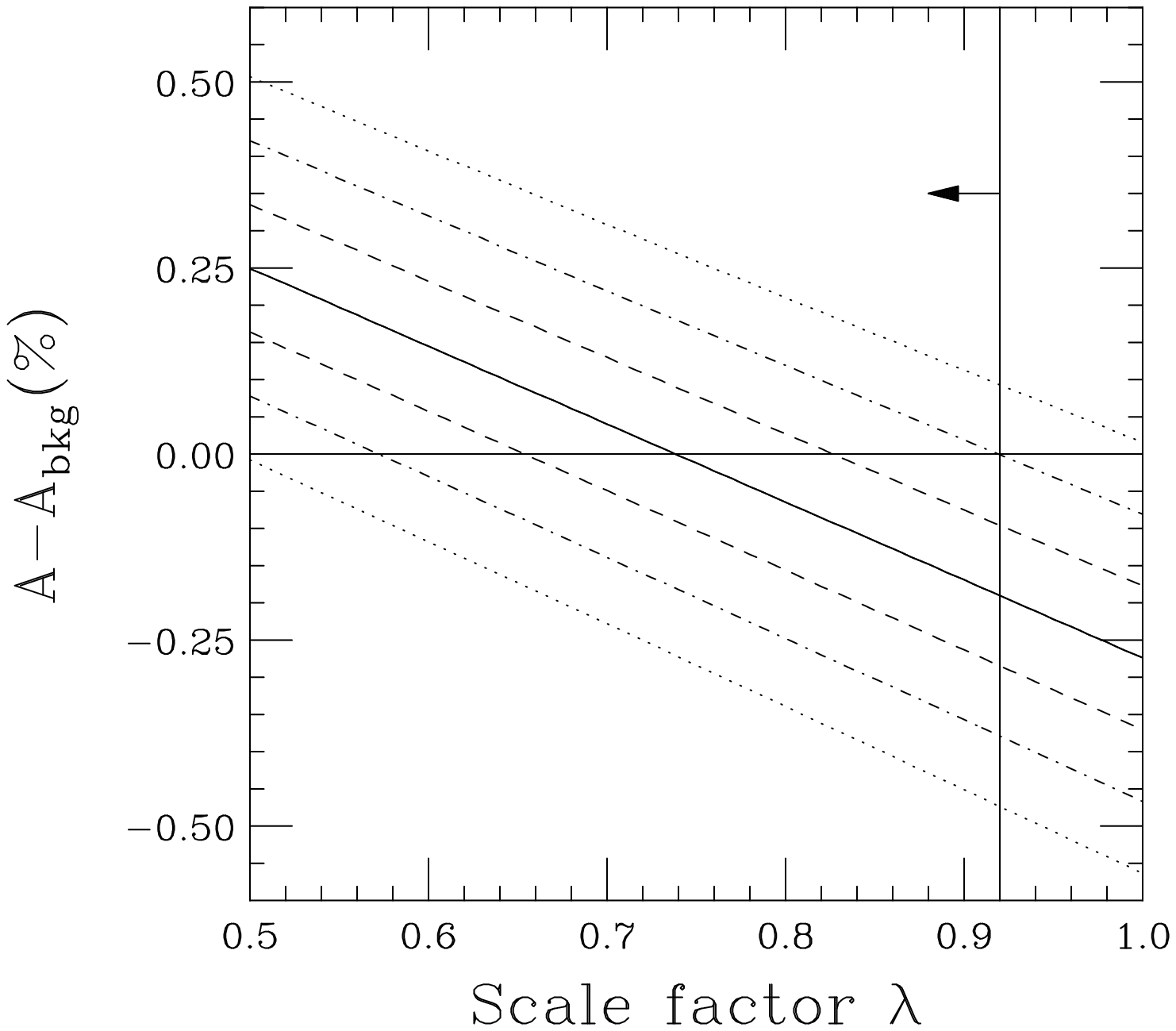}
\includegraphics[width=0.48\textwidth]{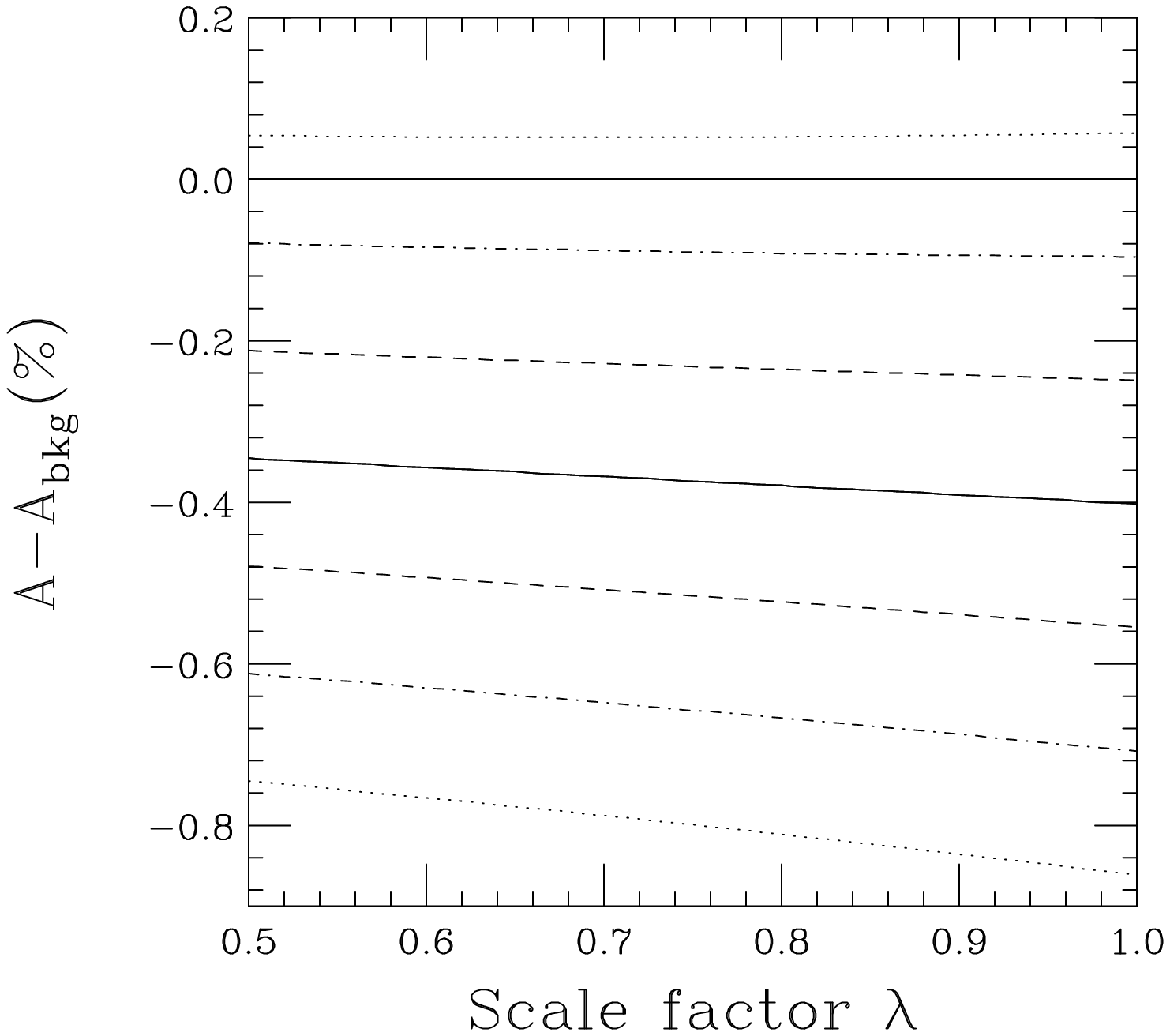}
\caption{Dependence of net dimuon asymmetry $A - A_{\rm bg}$ on scale factor
$\lambda$ for (left) IP $< 120~\mu$m; (right) IP $> 120~\mu$m.  Lines as in
Fig.\ \ref{fig:nocut}.  $A - A_{\rm bg}$ is within $2 \sigma$ of zero for
$\lambda < 0.92$ (left) but for no positive value of $\lambda$ when
IP $> 120~\mu$m.}
\end{figure}

The effect of rescaling the contribution of $F_K A_K$ may be seen in
Figs.\ 1 - 4.  For no constraint on IP and for
{\it maximum} values of IP $< (50,80,120)~\mu$m, a common
choice of $\lambda \simeq 2/3$ shifts derived values of $A - A_{\rm bkg}$ to
within $1 \sigma$ of zero.  With the exception of IP $> 120~\mu$m, it is always
possible to choose a value of $\lambda > 0.66$ such that $A - A_{\rm bg}$ is
within $2 \sigma$ of zero.  However, for IP $> 120~\mu$m, with $\lambda = 1$
the value of $A - A_{\rm bkg} = -0.402 \pm 0.152$ is $2.6 \sigma$ from zero,
and quite insensitive to rescaling of $F_K A_K$.

Effects of rescaling the kaon asymmetry background may also be studied in the 
inclusive muon asymmetry $a$,
which is less sensitive than $A$ to a nonzero value of $A^b_{sl}$.
The measured values of $a - a_{\rm bkg}$ for all IP constraints are consistent 
with zero~\cite{Abazov:2011yk,D0web}. This is expected because the contribution 
of $A^b_{sl}$ to this difference is proportional to $c_b$ which is much smaller 
than $C_b$. (See Eq.~(\ref{A}) above and Eq.~(7)  in 
Ref.~\cite{Abazov:2011yk} with values of $C_b$ and $c_b$
given in this reference.) We checked that rescaling the kaon background 
contribution to $a$ by $\lambda \simeq 2/3$ leaves values of $a- a_{\rm bkg}$ 
consistent with zero within $1\sigma$ for IP $ > 80, 120~\mu$m and within less than 
3$\sigma$ for IP $ > 50~\mu$m. However, for IP $< 50, 80, 120~\mu$m this scaling 
moves $a - a_{\rm bkg}$ significantly (6$\sigma$) away from zero.
Thus we conclude that a solution $A^b_{sl} =0$ cannot be obtained consistently for
all IP constraints by introducing a common rescaling factor for the background 
kaon asymmetry. 

\section{MAXIMUM \& MINIMUM IP FOR TWO $\mu$'S}
.\label{sec:othercuts}

So far D0 has analyzed dimuon events in which the two muons have IP either 
less or greater than the values 50, 80 and 120 $\mu$m. Other events, which 
have not yet been studied in the D0 sample, involve one muon with IP 
less than 50, 80 or 120 $\mu$m and a second muon having IP greater than the 
same value. The numbers of events in the three classes,
$N^{<<}_{\mu\mu}, N^{>>}_{\mu\mu}$ and $N^{<>}_{\mu\mu}$, are listed in 
Table \ref{tab:3classes} for the three IP values. This Table also quotes the calculated 
total number of dimuon events $N^{\rm total}_{\mu\mu}$ for the three IP values. 
The total numbers are in reasonable 
agreement with each other. Their deviation by $7\%$ relative to
the number $6.019\times 10^6$ quoted in Ref.\ \cite{Abazov:2011yk} for events
with no IP constraint may be due to a second order effect~\cite{GBPC}.

\begin{table}[h]
\caption{Number of dimuon events with muons having IP either below or above 
50, 80, and 120 $\mu$m. 
\label{tab:3classes}}
\begin{center}
\begin{tabular}{c c c c c} \hline \hline
IP & $N^{<<}_{\mu \mu}$ & $N^{>>}_{\mu\mu}$ & $N^{<>}_{\mu\mu}$ & 
$N^{\rm total}_{\mu\mu}$ \\
 \hline
  ($\mu$m)   &  $(10^6)$  &     $(10^6)$      &     $(10^6)$  &  $(10^6)$  \\
      50     &   1.527    & 1.680  & 3.203 & 6.410\\
      80     &    2.174   &  1.152 & 3.165 & 6.491 \\
     120     &    2.857  & 0.714  & 2.856 & 6.427 \\ \hline \hline
\end{tabular}
\end{center}
\end{table}

Noting that about half of the total like-sign dimuon events have not
yet been studied experimentally
($N^{<>}_{\mu\mu}/N^{\rm total}_{\mu\mu} \sim 1/2$), we wish to discuss 
their expected charge asymmetries.
For this discussion we need the fraction of background events $F^{<>}_x (x=K, \pi, p)$.
Using 
\beq
N^{<>}_{\mu\mu}=2(N^{<<}_{\mu\mu}N^{>>}_{\mu\mu})^{1/2}~,~~~
N^{<>}_x=2(N^{<<}_xN^{>>}_x)^{1/2}~,~~~
N_x = F_x N_{\mu\mu}~,
 \eeq
 one has
\beq
\label{F<>}
 F^{<>}_x = (F^{<<}_x\,F^{>>}_x)^{1/2}~,~~~~(x=K, \pi, p)~.
 \eeq
 Thus, fraction of background events from kaons, pions and protons in 
 the class $<>$ under consideration may be calculated from corresponding
 fractions  given in Tables XXI and XXII of Ref.\ \cite{Abazov:2011yk} for the two 
 classes $<<$ and $>>$. 
 These fractions are listed in Table \ref{tab:frac} for IP$=50, 80, 120~\mu$m.
 The three calculated fractions of kaons for events of the class $<>$ are equal 
 within 1$\sigma$ to the fraction measured by D0 for events with no restrictions on 
 IP~\cite{Abazov:2011yk} , $F_K\times 10^2=13.78 \pm 0.38$. 
 We note that  kaon fractions $F^{<<}_K$ are about twice larger than the fractions
 $F^{<>}_K$, while $F^{>>}_K$ are about twice smaller than these fractions. 
  
\begin{table}[h]
\caption{Fractions of background events of type 
$<<$ and $>>$ from Ref.\ \cite{Abazov:2011yk} and fractions 
of events of type $<>$ given by Eq.~(\ref{F<>}) for IP = 50, 80, 120~$\mu$m. 
\label{tab:frac}}
\begin{center}
\begin{tabular}{c c c c} \hline \hline
Fraction  & IP $= 50~\mu$m & IP $= 80~\mu$m & IP $= 120~\mu$m \\ \hline
 $F^{<<}_K \times 10^{-2}$     & 28.03 $\pm$ 0.95 & 23.79 $\pm$ 0.74 & 
 21.49 $\pm$ 0.62\\ 
 $F^{<<}_\pi \times 10^{-2}$     & 51.72 $\pm$ 3.18 & 44.26 $\pm$ 2.63 
 & 40.47 $\pm$ 2.26 \\   
 $F^{<<}_p \times 10^{-2}$     & 0.77 $\pm$ 0.29 & 0.66 $\pm$ 0.25 & 
 0.59 $\pm$ 0.23\\ 
  $F^{>>}_K \times 10^{-2}$     & 6.31 $\pm$ 1.73 & 4.79 $\pm$ 2.59 & 
 4.48 $\pm$ 4.05\\ 
 $F^{>>}_\pi \times 10^{-2}$     & 9.51 $\pm$ 2.36 & 6.39 $\pm$ 2.95 
 & 4.43 $\pm$3.95 \\   
 $F^{>>}_p \times 10^{-2}$     & 0.11 $\pm$ 0.06 & 0.03 $\pm$ 0.04 & 
 0.03 $\pm$ 0.05\\  \hline
 $F^{<>}_K \times 10^{-2}$     & 13.30 $\pm$ 1.84 & 10.67 $\pm$ 2.89 & 
 9.81 $\pm$ 4.44\\ 
 $F^{<>}_\pi \times 10^{-2}$     & 22.18 $\pm$ 2.84 & 16.82 $\pm$ 3.91 
 & 13.39 $\pm$ 5.98 \\   
 $F^{<>}_p \times 10^{-2}$     & 0.29 $\pm$ 0.10 & 0.14 $\pm$ 0.10 & 
 0.13 $\pm$ 0.11\\   \hline \hline
\end{tabular}
\end{center}
\end{table}

Results of asymmetries for background sources are given in Table
\ref{tab:asym}.  
The calculated background asymmetries, for samples involving one
muon with IP larger than 50, 80 or 120 $\mu$m and a second muon with IP smaller
than the same value, should be compared with corresponding future results by
D0.  They may be used to extract the net charge asymmetry for $B_{(s)}$ decays 
$A^b_{sl}$  from the raw asymmetry $A$.
 
\begin{table}[h]
\caption{Asymmetries calculated for background events of type 
$<>$ defined in the text for IP = 50, 80, 120 $\mu$m. 
\label{tab:asym}}
\begin{center}
\begin{tabular}{c c c c} \hline \hline
Asymmetry & IP $= 50~\mu$m & IP $= 80~\mu$m & IP $= 120~\mu$m \\ \hline
$F^{<>}_K A_K \times 10^2$                  & $+0.649 \pm 0.091$ 
 & $+0.521 \pm 0.141$ & $+0.479 \pm 0.217$ \\
$F^{<>}_\pi A_\pi \times 10^2$              & $-0.007 \pm 0.018$
 & $-0.005 \pm 0.013$ & $-0.004 \pm 0.011$ \\
$F^{<>}_p A_p \times 10^2$                  & $-0.002 \pm 0.010$
 & $-0.001 \pm 0.005$ & $-0.001 \pm 0.005$ \\
$(2 - F^{<>}_{\rm bkg}) \Delta \times 10^2$ & $-0.217 \pm 0.032$
 & $-0.228 \pm 0.033$ & $-0.233 \pm 0.035$ \\ \hline
$A^{<>}_{\rm bkg} \times 10^2$              & $+0.423 \pm 0.099$
 & $+0.287 \pm 0.145$ & $+0.241 \pm 0.220$ \\  
\hline \hline
\end{tabular}
\end{center}
\end{table}

\section{CONCLUSIONS}
\label{sec:conc}

We have examined the relation between the impact parameter (IP) selections
performed in Ref.\ \cite{Abazov:2011yk} and those we suggested in Ref.\
\cite{Gronau:2010cw}.  
We find that the minimum transverse IP of (50, 80, 120) $\mu$m
considered in Ref.\ \cite{Abazov:2011yk} is equivalent to a value of the
parameter $b_0$ which exceeds IP by factors of 2.7 to 3.6.  For the most
stringent criterion, taking events with IP $< 50~\mu$m for both muons the sample
of dimuons from $B_{(s)}$ decays is reduced to about 1/6 the size of the sample 
involving no bounds on IP. Fractions of backgrounds from kaons and pions in 
the former sample are each
about twice as large as in the latter sample. In spite of the considerably smaller 
sample of signal events with IP $<$ 50 $\mu$m and the larger background, 
the statistical significance of  $A - A_{\rm bkg}=(-0.527 \pm 0.127)\%$ measured 
for IP $<$ 50 $\mu$m is $4.1\sigma$, the same as measured for dimuons with
no IP constraint, $A - A_{\rm bkg} = (-0.276 \pm 0.067)\%$. (See Tables III and IV.)

We asked whether a rescaling of the parameter $F_K A_K$ describing kaon
background asymmetry could lead to annulment of the claimed charge asymmetry 
for same-sign muon pairs from $B_{(s)}$ decays.
For selections of {\it maximum} impact parameters IP $< (50,80,120)~\mu$m, a 
rescaling by a factor of $\lambda \simeq 2/3$ led to reduction of the net
charge asymmetry to within $1 \sigma$ of zero.  However, for
{\it minimum} impact parameters IP $> (50,80,120)$
$\mu$m, with greatly reduced
kaon backgrounds, this was not so, and for IP $> 120~\mu$m, no positive choice
of $\lambda$ reduced the net asymmetry to less than $2 \sigma$ from zero.
In contrast, while introducing a rescaling factor $\lambda \simeq 2/3$ in inclusive 
muon samples is consistent with $A^b_{sl}=0$ for IP $ > 50, 80, 120~\mu$m, it 
leads to a nonzero asymmetry for IP $< 50, 80, 120~\mu$m. 

We calculated background asymmetries for dimuon samples in which one muon
has a maximum IP while the other has a minimum IP, for the three cases
IP = 50, 80, 120 $\mu$m. These calculated asymmetries, expected to be confirmed in
future studies by D0, may be used for measuring $A^b_{sl}$ in these samples. 

The fraction of dimuons from  kaons was seen to decrease by imposing 
a minimim value for the impact parameter IP. 
For instance, the background asymmetry from kaons was reduced from 
$F_KA_K= (0.633 \pm 0.031)\%$ with no IP restriction to 
$F_KA_K=(0.205\pm 0.060)\%$ for IP $>$ 50 $\mu$m. The corresponding
number of like-sign dimuon events decreased by about a factor $1/4$
(see Table IX), and the significance of a nonzero $A - A_{\rm bkg}$ went
down from $4.1\sigma$ to $2.4\sigma$ (see Table V). 
In future studies of the same-sign dimuon charge asymmetry we thus advocate
emphasis on reduction of kaon background by choosing a minimum value of
impact parameter, even if at the cost of statistics.  Such studies could, in
principle, be performed by other collaborations such as CDF at the Fermilab
Tevatron and LHCb at the CERN Large Hadron Collider.

\section*{Acknowledgments}

We thank G. Borissov and B. Hoeneisen for 
very useful discussions, and the Aspen
Center for Physics for congenial hospitality during part of this study.
This work was supported in part by the United States Department
of Energy through Grant No.\ DE FG02 90ER40560.

\end{document}